\documentclass[preprint]{elsarticle}
        \usepackage{amsmath}
        \usepackage{amssymb}
        \usepackage{textcomp}
        \usepackage{stmaryrd}
        \usepackage{makeidx}
        \usepackage{amsfonts}
        \usepackage{graphicx}
        \usepackage{dblfloatfix}
        \usepackage[ansinew]{inputenc}
        \usepackage{geometry}
        \usepackage{natbib}        
        \usepackage{subfigure}
        \usepackage{epsfig}
        \usepackage{color}
        \usepackage[colorlinks,pagebackref,hyperindex]{hyperref}
        \usepackage{setspace}
        \hypersetup
        {
                colorlinks,%
                citecolor=black,%
        	      linkcolor=black,%
                urlcolor=black,%
        }

\makeindex

\begin{document}

\title{Scanning tunneling microscopy at multiple voltage biases of stable ``ring-like'' Ag clusters on Si(111)--(7$\times$7)}

\author[aebi]{N. Mariotti\corref{cor1}}
\ead{nicolas.mariotti@unifr.ch}

\author[aebi]{C. Didiot}
\author[aebi]{E.F. Schwier}
\author[psi]{C. Monney}
\author[pv]{L.-E. Perret-Aebi}
\author[pv]{C. Battaglia}
\author[aebi]{M.G. Garnier}
\author[aebi]{P. Aebi}

\address[aebi]{D\'epartement de physique, Universit\'e de Fribourg and Fribourg Center for Nanomaterials\\ Chemin du Mus\'ee 3, 1700 Fribourg, Switzerland}

\address[psi]{Swiss Light Source, Paul Scherrer Institut, 5232 Villigen PSI, Switzerland}

\address[pv]{\'Ecole Polytechnique F\'ed\'erale de Lausanne (EPFL), Institute of Microengineering (IMT), Photovoltaics and Thin Film Electronics Laboratory, 2000 Neuchâtel, Switzerland}

\cortext[cor1]{Corresponding author}

\date{\today}

\begin{abstract}
Since more than twenty years it is known that deposition of Ag onto Si(111)--(7$\times$7) leads under certain conditions to the formation of so-called ``ring-like'' clusters, that are particularly stable among small clusters. In order to resolve their still unknown atomic structure, we performed voltage dependent scanning tunneling microscopy (STM) measurements providing interesting information about the electronic properties of clusters which are linked with their atomic structure. Based on a structural model of Au cluster on Si(111)--(7$\times$7) and our STM images, we propose an atomic arrangement for the two most stable Ag ``ring-like'' clusters.
\end{abstract}


\maketitle

\section{Introduction}
Self-organized nanostructures at surfaces have been intensively studied for a wide variety of systems \cite{Brune,Barth,selfCysteine,Sun,metalChainsOnPt}. Interest in this field is mainly driven by the technological need for miniaturization \cite{Barth,CoAu111}, as nanopatterned surfaces are one of the main means to direct the growth of nanostructures. In particular, the Si(111)--(7$\times$7) surface reconstruction is a very good template for the growth of self-organized arrays of metallic nanoclusters. Indeed, its large unit cell and the high barrier for the hopping of deposited metallic atoms between the two half unit cells allow for the formation of atomically precise clusters\cite{AlArraySi,AlArraySiJia,GaArraySi,GaArraySiPRL,InArraySi}. 

Growth of Ag on Si(111)--(7$\times$7) has been extensively studied for more than twenty years \cite{ToschNeddermeyer}. Preferential nucleation inside faulted half-unit cells (FHUCs) for Ag clusters on Si(111)--(7$\times$7) has been demonstrated for a wide range of deposition rates by O\v st'\'adal et \emph{al.}\cite{NucleationAllRates}. In the case of low Ag coverage, they showed that a sufficiently strong annealing eliminates most of the less stable clusters located in FHUCs thereby promoting homogeneity \cite{annealingOstadal}. 

Some of the most stable clusters appear as ``ring-like'' structures when imaged by means of STM. This structural stability makes them particularly interesting for the study of self-organized arrays of well-defined Ag nanoclusters. Moreover, despite the fact that their electronic properties were studied recently\cite{elecStructAgOn7x7}, the geometrical structure of these ``ring-like'' clusters is still unknown. 

Here, we have grown self-organized arrays of Ag clusters on the Si(111)--(7$\times$7) surface reconstruction where Ag deposition was followed by annealing. We present a series of STM images at different bias voltages for the two different ``ring-like'' clusters. In order to propose a structural model for these clusters, we combined this new insight together with the structure of Au clusters on Si(111)--(7$\times$7) studied by Ghose et \emph{al.}\cite{AuSXRD} as well as the number of Ag atoms determined  by comparison with the studies from Ming et \emph{al.}\cite{mingXiao,AgAssDesassembly}.

\section{Experiment}
Phosphorus n-doped (111)-oriented Si crystals with a room temperature resistivity between 0.001$-$0.005 $\Omega$.cm were used as substrates. Our experiment was carried out with an Omicron low temperature scanning tunneling microscope (LT-STM) operated in ultrahigh vacuum with a base pressure better than $6\cdot 10^{-11}$ mbar. All measurements presented in this paper were performed at 77~K with Pt-Ir tips. The Si(111)--(7$\times$7) surface reconstruction was obtained by direct current resistive heating. 
First, the sample was heated up to 1300~K with a base pressure below $5\cdot 10^{-10}$ mbar, then it was flashed repeatedly over 1500~K to clean the surface and then slowly cooled down across the (7$\times$7) transition range of temperature (around 1130~K\cite{T7x7}). 
Typically, more than 10 hours are spent to ensure that the sample remains enough time close to the transition temperature\cite{Corsin331}. Sample temperature was monitored using a pyrometer. Ag was deposited with an e-beam evaporation cell and the pressure during evaporation never exceeded $2\cdot 10^{-10}$ mbar. Surface quality was checked using low energy electron diffraction and STM.

\begin{figure}[!h]
	\centering
		\includegraphics[width=0.5\textwidth]{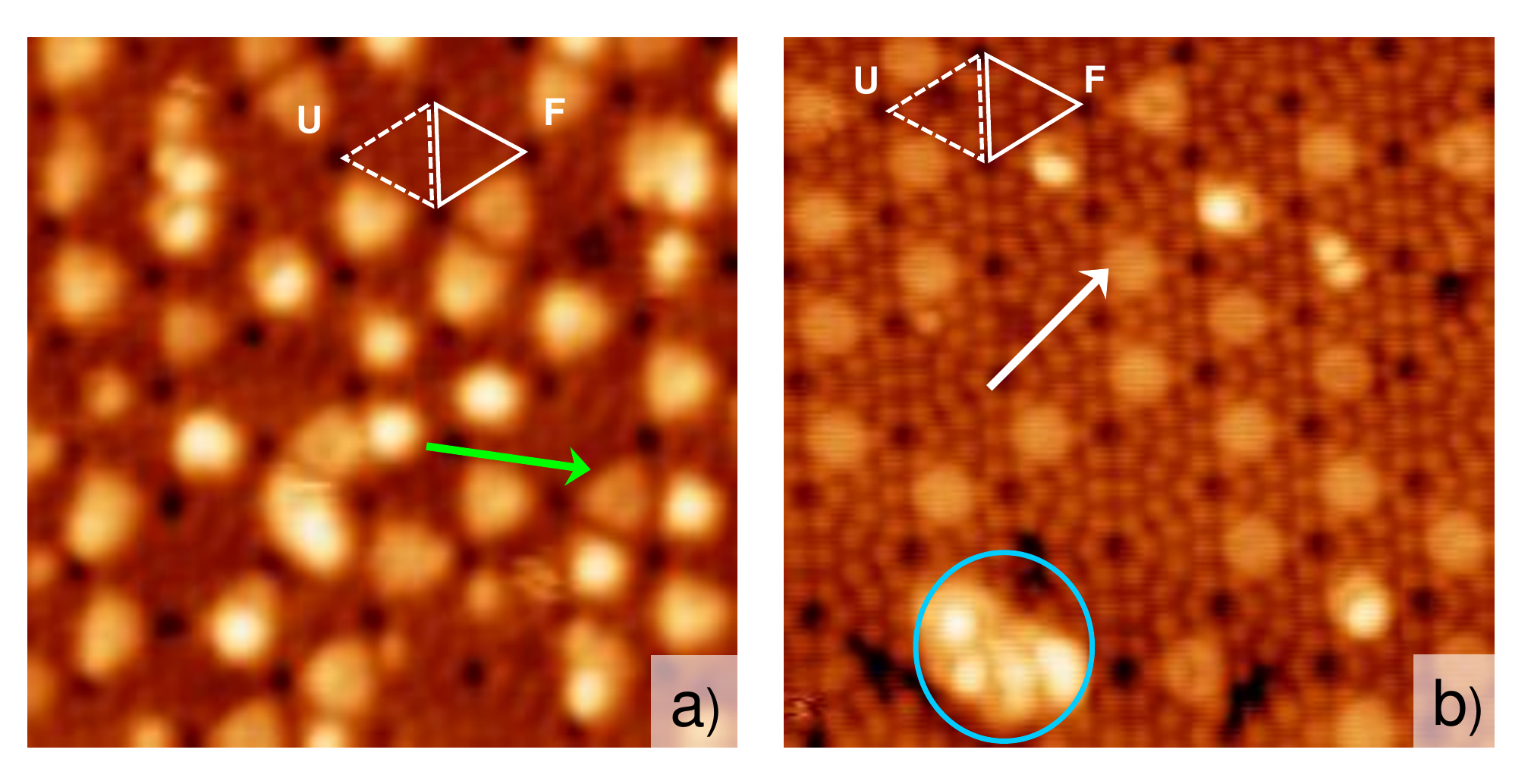}
	\caption{a) STM image of Ag deposited at 420~K on Si(111)--(7$\times$7) surface. The green arrow indicates a cluster inside a UHUC. b) The same sample after annealing at 620~K, the white arrow indicates a typical stable cluster in a FHUC, the blue circle a coalesced cluster.  V$\mathrm{_{bias}}$=+1.5~V, I$\mathrm{_{set}}$=0.2 nA, 16$\times$16 nm$^{2}$}
	\label{fig:fig1}
\end{figure}

To optimize the growth of arrays of well-defined clusters we have varied the deposition rate, the substrate temperature during evaporation, the total Ag coverage and the annealing temperature. Based on the work of Koc\'an et \emph{al.}\cite{PrefVsRate} and preliminary measurements, we have established that the deposition rate should be minimized to increase preference for FHUC and overall order. The deposition rate was $4 \cdot 10^{-5}~ 1\mathrm{ML} \cdot \mathrm{s}^{-1}$ ($\mathrm{ML}$ here corresponds to $6.7 ~ \mathrm{atoms} / \mathrm{nm}^{2} = 6.7 \cdot 10^{14} ~ \mathrm{atoms} / \mathrm{cm}^{2}$). The flux was determined with an STM measurement of the total fractional coverage of the post-annealing Si(111)--($\sqrt{3}\times \sqrt{3}$)--Ag surface~\cite{r3r3Wilson}. We checked the steadiness of the flux by Ag depositions of various durations on the Si(111)--(7$\times$7) surface and subsequent STM measurements. The given deposition rate corresponds approximately to the deposition of one Ag atom per FHUC in  10 minutes. The substrate temperature was kept close to 420~K during deposition to minimize the growth of cluster inside the unfaulted half-unit cell (UHUC), without inducing interdiffusion between Ag adatoms and the Si substrate. A coverage of 0.25~ML was chosen in order to maximize the occupation of the FHUCs with stable clusters. 
By annealing, less stable clusters tend to dissolve, leading to a redistribution of Ag atoms that strongly favors clusters that contain between 6 and 18 atoms, hence increasing homogeneity\cite{annealingOstadal}. The effect is illustrated in Fig.~\ref{fig:fig1}, which shows the typical result of an annealing just below the temperature of the $\sqrt{3}\times\sqrt{3}$ transition. 

In Fig.~\ref{fig:fig1} a), we present a topographic STM image of a typical sample before annealing. A preferential growth in the FHUCs is observed, but nucleation in the UHUCs (green arrow) is not negligible. Also, there is a large variation in the shapes of the clusters inside FHUCs. In Fig.~\ref{fig:fig1} b), an annealing was performed on the same sample. The majority of nanostructures are now well-defined ``ring-like'' Ag clusters (white arrow), accompanied with some coalesced nanostructures (blue circle). The limit in ordering and cluster regularity is reached by annealing for half an hour at around 620~K.

\begin{figure}[!h]
	\centering
		\includegraphics[width=0.45\textwidth]{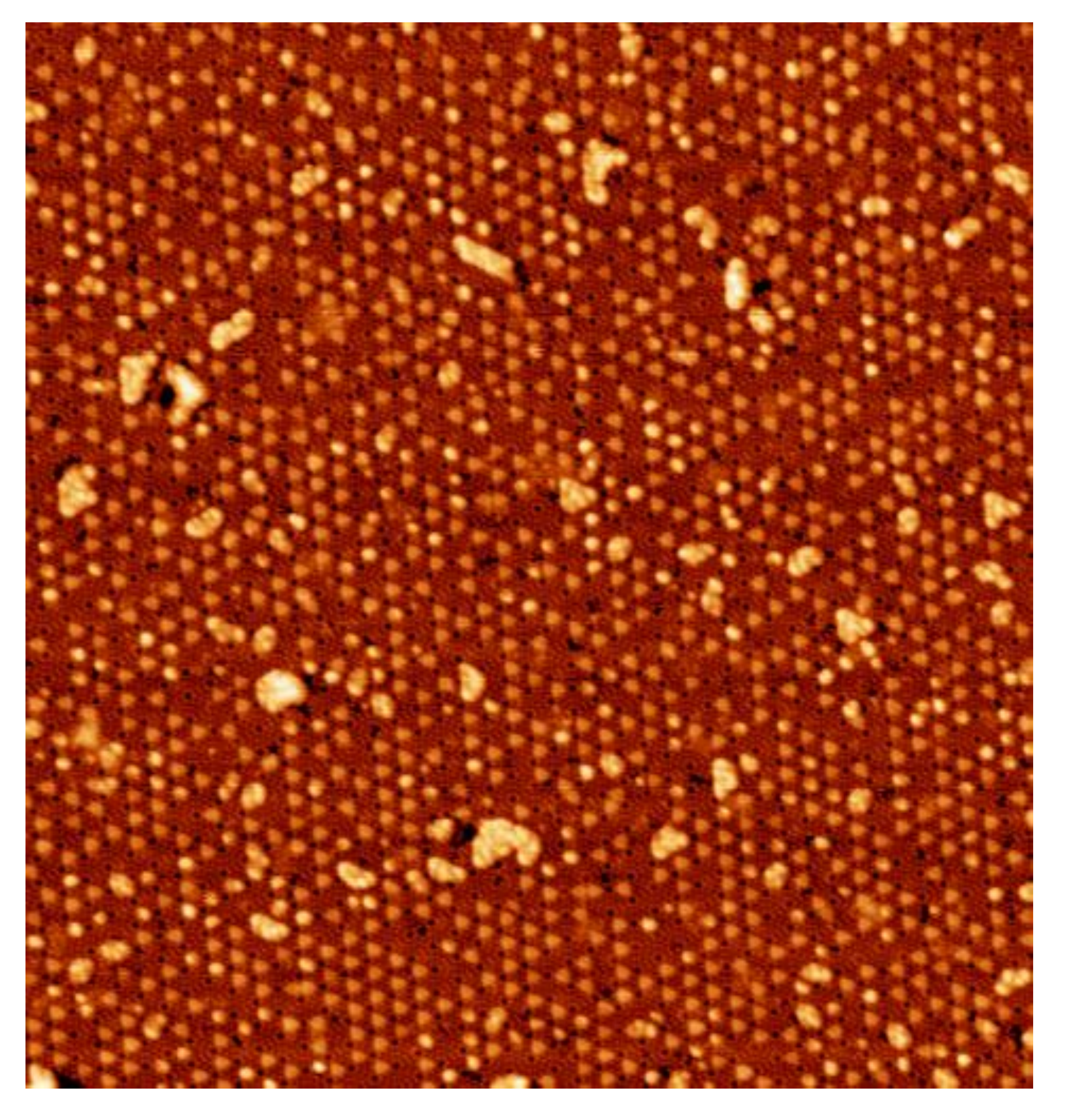}
	\caption{Large scale STM image of the 0.25 ML Ag on Si(111)--(7$\times$7) annealed at 620~K. V$\mathrm{_{bias}}$=+2~V, I$\mathrm{_{set}}$=0.1 nA, 100$\times$100 nm$^{2}$}
	\label{fig:fig2}
\end{figure}

To compare the quality of ordering with other systems, we have performed statistics on the basis of around 3000 half-unit cells (HUCs). Fig.~\ref{fig:fig2} shows one of the three large scale STM images of the Ag/Si(111)--(7$\times$7) system used for the analysis. One can observe small arrays of clusters that cover an important fraction of the surface. However empty FHUCs, clusters inside UHUCs and coalesced clusters are also present. 

First, about half of the FHUCs on the surface are covered with triangular clusters while the rest of the FHUCs are either empty, or covered with other types of clusters (e.g. coalesced clusters).
Second, 83 \% of the triangular clusters are located inside FHUCs (less than the 95 \% observed for Tl clusters\cite{TlArraySi}). Third, coalesced clusters cover 7\% of all the HUCs. Even if theses arrays are not as atomically precise as model systems such as Al and Tl on Si(111)--(7$\times$7) \cite{AlArraySi,AlArraySiJia,TlArraySi}, both homogeneity and long range order may be sufficient to allow for a study of this system by means of a space integrating measurement technique like photoemission.

\section{Structural models}

\begin{figure}[!h]
	\centering
		\includegraphics[width=0.45\textwidth]{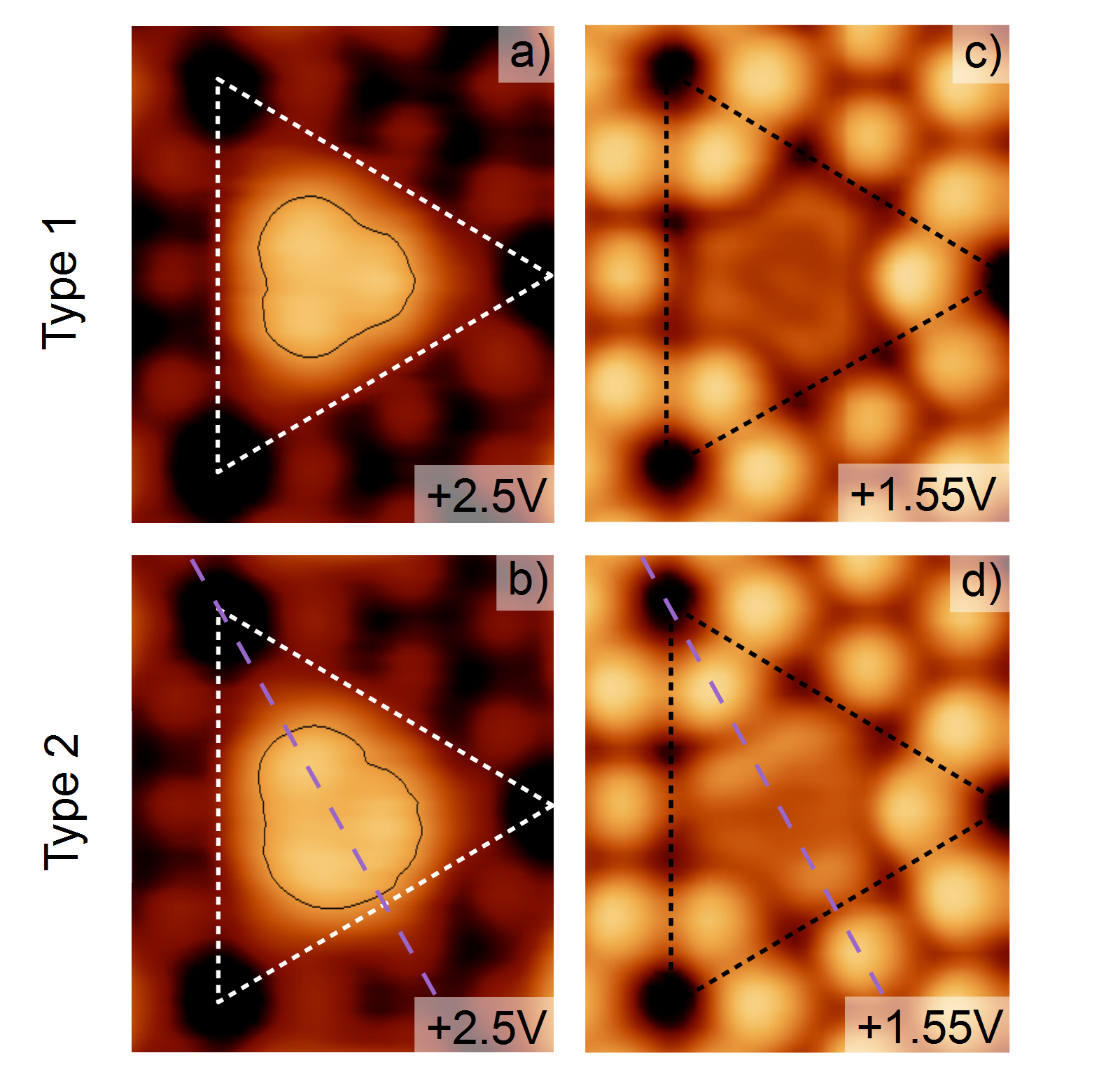}
	\caption{Empty-states STM images at two bias voltages illustrating the two types of stable ``ring-like'' Ag clusters. a), b) Clusters imaged at +2.5~V. Associating the whole color variation to the 0.28 nm corrugations of clusters and Si adatoms, and plotting one corrugation isoline enhances lobes visibility. c), d) Same clusters imaged at +1.55~V. The symmetry axis of the ``Type 2'' structure is indicated by a magenta dashed line; the FHUC is outlined by the dashed-line triangles. I$\mathrm{_{set}}$ = 0.4 nA, 3$\times$3.5 nm$^{2}$}
	\label{fig:fig3}
\end{figure}

In Fig.~\ref{fig:fig3}, we present STM images of two different types of triangular clusters which constitute the self-assembled arrays (Type 1 \& 2). At +2.5~eV above the Fermi level, both exhibit a three lobe structure, highlighted in  Fig.~\ref{fig:fig3} a) and b). A difference of symmetry is already visible and becomes even more pronounced at low bias voltages, as shown in Fig.~\ref{fig:fig3} c) and d). Type~1 exhibits a 3-fold symmetry while Type~2 demonstrates a symmetry along a mirror plane symbolized by the magenta dashed line. The symmetry difference can be explained most likely by different numbers of Ag atoms inside the two types of clusters. 

Ming et \emph{al.} recently determined the number of atoms in Ag clusters obtained by deposition at room temperature without annealing\cite{mingXiao,AgAssDesassembly}. While measuring at a bias voltage of +2~V, they observed three lobes only on clusters with 10 to 13 Ag atoms. Our STM images (Fig.~\ref{fig:fig3} a \& b) taken at +2.5~V closely resemble those clusters with 10 and 11 Ag atoms, respectively\cite{AgAssDesassembly}. It should be noted that in our case, measurements were performed at lower temperature. In the "3 lobes" range between 10 and 13 atoms,  Ag clusters with 10 and 11 atoms have a lifetime many times larger than clusters with 12 or 13 atoms\cite{mingXiao}. As a consequence, only Ag$\mathrm{_{10}}$ and Ag$\mathrm{_{11}}$ can be considered to remain after the annealing at 620~K.

In order to define the atomic positions of Ag inside both types of clusters, we use a comparative Ansatz to propose a simple structural model. Because Au lies in the same chemical group as Ag and has a comparable radius, similarities might exist between clusters. There exist two different models for Au clusters on Si(111)--(7$\times$7).
Wu et \emph{al.} have studied Au clusters by means of voltage dependent STM images and proposed a model based on calculations for Au clusters only containing 6 atoms\cite{AuClus}. The second model available was proposed by Ghose et \emph{al.} using 9 Au atoms grown inside FHUCs\cite{AuSXRD}. This structural model will be used as a basis for our model. It is interesting to note that six of the atom positions correspond to minima of the static potential energy for a single Ag atom calculated by Wang et \emph{al.}\cite{AgBasin}. Other positions do not correspond to the other energy minima. The location of the Ag trimers observed by Hu \emph{al.} \cite{rectifAgTrimer} makes this difference plausible (see Ref. \cite{rectifAgTrimer}).

\begin{figure}[!h]
	\centering
		\includegraphics[width=0.45\textwidth]{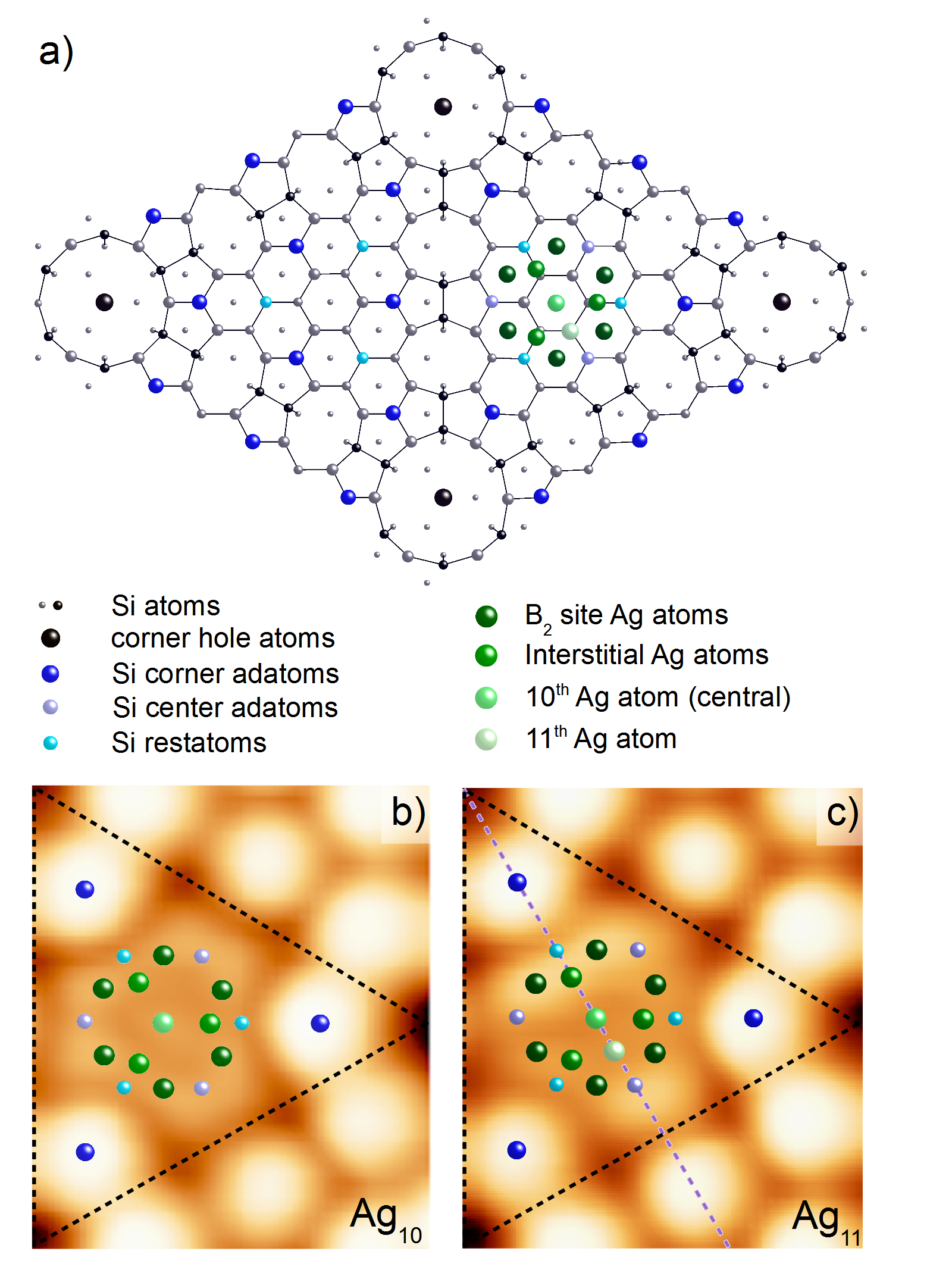}
	\caption{a) Schematic representation of the proposed structural model of Ag$_{10}$ and Ag$_{11}$ nanoclusters on the faulted HUC of the Si(111)--(7$\times$7) surface. b) and c) Ag$_{10}$ and Ag$_{11}$ cluster models superimposed on corresponding STM images. The magenta dashed line in c) represents the symmetry axis of Ag$_{11}$. $V\mathrm{_{bias}} = +1.55~V$; I$\mathrm{_{set}}$ = 0.4~nA, 2.4$\times$2.7~nm$^{2}$}
	\label{fig:fig4}
\end{figure}

Fig.~\ref{fig:fig4} a) shows a ball and stick model of the proposed positions. We propose that 9 Ag atoms occupy the same positions as the 9 Au atoms. As shown above, there are two types of clusters with different symmetry (3-fold and mirror plane) composed of either 10 or 11 Ag atoms. In a Ag$\mathrm{_{9+n}}$ cluster, it is not possible to put the 11$\mathrm{^{th}}$ Ag atoms without breaking the 3-fold symmetry. However, a 10$\mathrm{^{th}}$ Ag atom can be easily placed at the center of the FHUC which is also a high coordination site for adsorption (Fig.~\ref{fig:fig4} b). Also using a symmetry argument, we can estimate the position of the 11$\mathrm{^{th}}$ Ag atom. 
The mirror symmetry represented in the  Fig.~\ref{fig:fig4} c) implies that this additional Ag adatom must lie on the symmetry axis (magenta dashed line). 
The most likely location (very light green sphere) for it is between the 10$\mathrm{^{th}}$ Ag adatom and one of the three Si center adatoms (close to the on-top position of a Si atom in the first stacking fault layer). 

\begin{figure*}[h!]
	\centering
		\includegraphics[width=1\textwidth]{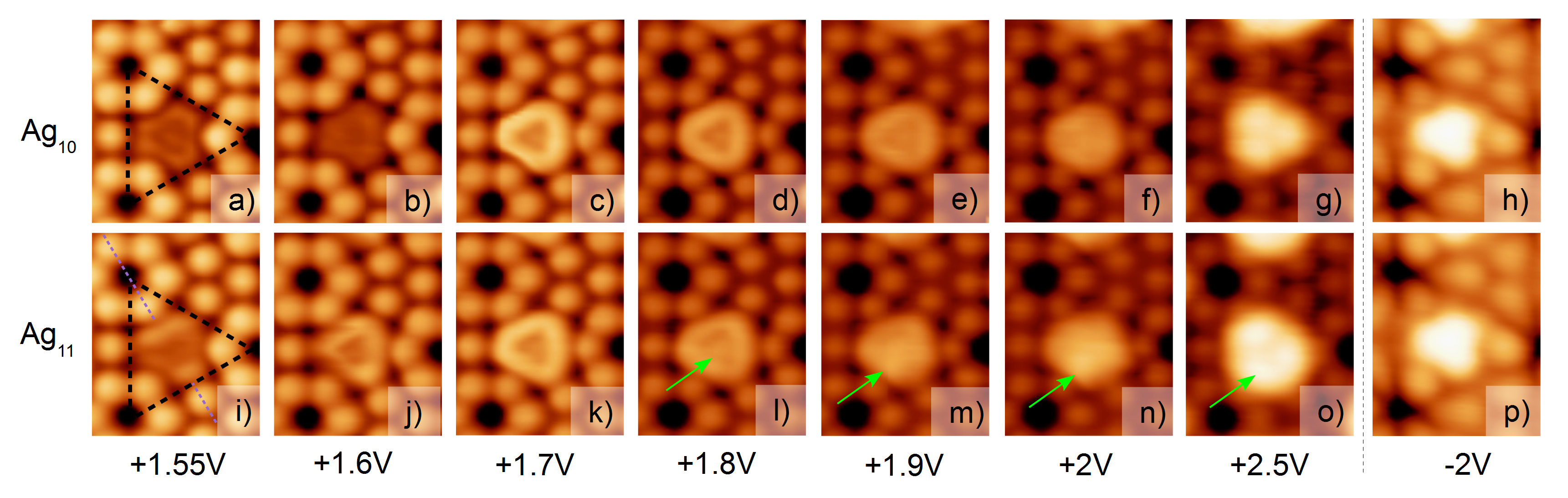}
	\caption{STM images at different bias voltages of respectively a Ag$_{10}$ cluster (a-h) and a Ag$_{11}$ cluster (i-p). The same lookup table was used for all STM images. I$\mathrm{_{set}}$ = 0.1 nA, 3$\times$4 nm$^{2}$}
	\label{fig:fig5}
\end{figure*}

The proposed models, shown in Fig.~\ref{fig:fig4}, cannot be directly compared with our apparent topographic STM images due to their strong dependence on the surface density of states. However voltage dependent STM images can provide spatial and energetic information about the electronic properties of clusters which are directly related to their structural properties.

\section{Voltage dependent STM images}
Two series of STM images obtained at eight different bias voltages for Ag$\mathrm{_{10}}$ and Ag$\mathrm{_{11}}$ are presented in Fig.~\ref{fig:fig5}. Empty-states (filled-states) STM images give us spatial and energetic information on unoccupied (occupied) electronic states within these two clusters. The Si corner adatoms of the FHUC are visible in all STM images below +2.5~eV and their apparent height close to the Ag cluster is similar to those of the bare Si(111)--(7$\times$7). This implies the absence of any substitution of Si with Ag at those sites\cite{Taniguchi}.

At 77~K and even with a highly-doped substrate, no STM image can be produced below +1.5~eV. Just above the onset of the conduction band, only the first unoccupied electronic state of the cluster can contribute to the tunneling current in the center of the FHUC (above the Ag cluster). In this case, STM can give us access to the local density of states. The first electronic state in the conduction bands can be observed at an energy close to +1.55~eV. For the Ag$\mathrm{_{10}}$ cluster, the density of states is clearly localized above the positions of the 10 Ag atoms proposed in our model and the 3 Si center adatoms (Fig.~\ref{fig:fig5} a \& \ref{fig:fig4} b). As already mentioned before, the additional atom in the case of Ag$\mathrm{_{11}}$ induces an asymmetrical change in the localization of the density of states with respect to Ag$\mathrm{_{10}}$ (Fig.~\ref{fig:fig5} i). This is made evident through a stronger localization visible as two bright areas perpendicular to the mirror plane.
Relatively to the 11$\mathrm{^{th}}$ Ag atom (see Fig.~\ref{fig:fig4} c), these maxima are respectively located near the closest Si center adatom and near the farthest three Ag adatoms (B$\mathrm{_{2}}$ site Ag atoms and interstitial Ag atom, see Fig.~\ref{fig:fig4}).

By increasing the bias voltage, additional electronic states start to contribute to the tunneling current and the well-known ``ring-like'' shape\cite{ToschNeddermeyer,elecStructAgOn7x7} of the clusters can be clearly observed at +1.7~eV (Fig.~\ref{fig:fig5} c \& k). 

In Fig.~\ref{fig:fig5} g), the density of states of Ag$\mathrm{_{10}}$ can be described with three maxima localized at the three Si restatom positions. This electronic eigenstate presents a maximum at +2.5~eV with a magnitude much larger than the lower lying electronic states within the clusters. There its contribution to STM measurements becomes predominant and is responsible for the apparent shape inversion of these triangular clusters. An additional contribution in the density of states of Ag$\mathrm{_{11}}$ starts to appear above +1.8~eV (Fig.~\ref{fig:fig5} l) at the proposed position of the 11$\mathrm{^{th}}$ Ag atom (Fig.~\ref{fig:fig4} c). This contribution is responsible for the obvious asymmetry visible up to +2.5~eV and indicated by green arrows on Fig.~\ref{fig:fig5} l -- o).

In Fig.~\ref{fig:fig5} h) \& p) occupied states were probed at $-$2~eV. For both clusters, the density of states is localized at the positions of the three center Si adatoms, leading to an inverted triangular appearance. The Si corner adatoms inside the FHUC are still visible, and the edges of the clusters are slightly bent towards the FHUC's center. The inverted triangular shape is similar for a wide range of negative bias voltages (from $-$1.5~eV to $-$3~eV) and well represented by Fig.~\ref{fig:fig5} h) \& p). This appearance was also observed on many systems such as Au, Zn, and Mn clusters on Si(111)--(7$\times$7)\cite{AuClus,ZnFilled,MnFilled}.

Furthermore, we can also observe some modifications of the electronic properties inside the bare UHUCs. While empty-states STM images show the six Si corner and center adatoms as on the bare Si(111)--(7$\times$7) (Fig.~\ref{fig:fig5} f \& n), three main protrusions are localized above the Si restatoms in filled-states STM images (Fig.~\ref{fig:fig5} h \& p). As already proposed by Li et \emph{al.} for Pb clusters on Si(111)--(7$\times$7)\cite{Pb_UHUC}, this modification of filled states in the bare UHUC could be induced by the nearby clusters in FHUCs.

\section{Conclusion}
We have grown well-ordered self-organized arrays of Ag clusters on Si(111)--(7$\times$7). At specific bias voltages, these nanostructures appear as the so-called ``ring like'' clusters. We were able to distinguish two types of clusters with different symmetries. Based on a model of Ghose et \emph{al.} for Au clusters, we proposed atomic structures for both types of clusters. Voltage dependent STM images were taken, giving us indirect information about the electronic structure of the clusters which may, in turn, be used to confirm first principles total energy calculations based on our structural model.

\section*{Acknowledgements}
\addcontentsline{toc}{section}{Acknowledgements}

Skillful technical assistance was provided by our workshop and electronic engineering team. STM image processing was done with the WSxM software\cite{WSxM}. This work was supported by the Fonds National Suisse pour la Recherche Scientifique through Div. II and the Swiss National Center of Competence in Research MaNEP.


\end{document}